\documentclass[prd,aps,showpacs,nofootinbib,preprint,eqsecnum]{revtex4-1}
\usepackage{graphicx,color,amsmath,amsxtra}
\usepackage{epsf}
\usepackage{amssymb}
\usepackage{enumerate}
\usepackage{hhline}
\usepackage{array}
\usepackage{tabularx}

\newcommand{\bear}{\begin{array}}  \newcommand{\eear}{\end{array}}
\newcommand{\bea}{\begin{eqnarray}}  \newcommand{\eea}{\end{eqnarray}}
\newcommand{\beq}{\begin{equation}}  \newcommand{\eeq}{\end{equation}}
\newcommand{\bef}{\begin{figure}}  \newcommand{\eef}{\end{figure}}
\newcommand{\bec}{\begin{center}}  \newcommand{\eec}{\end{center}}

  \newcommand{\abs}[1]{\vert{#1}\vert}

\newcommand{\Eqn}[1]{&\hspace{-0.2em}#1\hspace{-0.2em}&}

\def\be{\begin{equation}}
\def\ee{\end{equation}}
\def\bea{\begin{eqnarray}}
\def\eea{\end{eqnarray}}
\def\beq{\begin{eqnarray}}
\def\eeq{\end{eqnarray}}

\def\be{\begin{equation}}
\def\ee{\end{equation}}
\def\bea{\begin{eqnarray}}
\def\eea{\end{eqnarray}}
\def\beq{\begin{eqnarray}}
\def\eeq{\end{eqnarray}}

\begin{document}

\title{Thermodynamics of cosmological horizons in $f(T)$ gravity
}

\author{Kazuharu Bamba$^{1, 2}$\footnote{E-mail address: 
bamba@kmi.nagoya-u.ac.jp} and 
Chao-Qiang Geng$^{1,3}$\footnote{E-mail address: geng@phys.nthu.edu.tw} 
}
\affiliation{
$^1$Department of Physics, National Tsing Hua University, Hsinchu, Taiwan 300 
\\ 
$^2$Kobayashi-Maskawa Institute for the Origin of Particles and the
Universe,
Nagoya University, Nagoya 464-8602, Japan
\\
$^3$Physics Division, National Center for Theoretical Sciences, Hsinchu, Taiwan 300
}


\begin{abstract}

We explore thermodynamics of the apparent horizon in $f(T)$ gravity
with both equilibrium and non-equilibrium descriptions. 
We find the same dual equilibrium/non-equilibrium formulation for $f(T)$ 
as for $f(R)$ gravity. 
In particular, we show 
that the second law of thermodynamics can be satisfied 
for the universe with the same temperature outside and inside 
the apparent horizon. 

\end{abstract}



\maketitle

\section{Introduction}

The current accelerating expansion of the universe has been supported by many
 cosmological observations such as 
Type Ia Supernovae~\cite{SN1}, 
cosmic microwave background (CMB) radiation~\cite{WMAP, Komatsu:2010fb}, 
large scale structure~\cite{LSS}, 
baryon acoustic oscillations~\cite{BAO}, 
and weak lensing~\cite{WL}.
There are two representative 
approaches to explain the late time acceleration 
of the universe: 
One is to introduce some unknown matters called ``dark energy'' 
in the framework of general relativity 
(for a review on dark energy, see, e.g.,~\cite{Copeland:2006wr, Li:2011sd}). 
The other is to modify the gravitational theory, 
e.g., $f(R)$ gravity~\cite{Review-N-O, Sotiriou:2008rp, 
DeFelice:2010aj, Clifton:2011jh}. 

To explore gravity beyond general relativity, 
``teleparallelism" could be considered by using 
the Weitzenb\"{o}ck connection, which has no curvature but 
torsion, rather than the curvature defined by the Levi-Civita connection 
(see, e.g.,~\cite{Hehl:1976kj}).
This approach was also taken by Einstein~\cite{Einstein}. 
The teleparallel Lagrangian density described by the torsion scalar $T$ 
has been promoted to a function of $T$, $i.e.$, $f(T)$, 
in order to account for the late time cosmic 
acceleration~\cite{Bengochea:2008gz, Linder:2010py} 
as well as inflation~\cite{Inflation-F-F}. 
This concept is similar to the idea of $f(R)$ gravity.
Various aspects of 
$f(T)$ gravity have been examined in the 
literature~\cite{f(T)-Refs, 
BGL-Comment, 
Zheng:2010am,Local-Lorentz-invariance}.
In particular, the presence of extra degrees of freedom and the violation of local Lorentz invariance as well 
as the existence of non-trivial frames for $f(T)$ gravity have been noted~\cite{Local-Lorentz-invariance}.
Clearly, more studies on $f(T)$ gravity are needed to see if the theory is a viable one. 
Recently, the first law of thermodynamics in $f(T)$ gravity has been studied 
in Ref.~\cite{Miao:2011ki}\footnote{We note that the procedure for 
the discussions on the first law of thermodynamics in $f(T)$ gravity used 
in Ref.~\cite{Miao:2011ki} is different from that in this paper.}. 
In this paper, we concentrate on thermodynamics in $f(T)$ gravity. 

Black hole thermodynamics~\cite{BCH-B-H}
indicated 
the fundamental connection between gravitation and thermodynamics 
(for recent reviews, see~\cite{Rev-T-G-Pad}). 
In Ref.~\cite{Jacobson:1995ab}, 
the Einstein equation was derived from 
the Clausius relation in thermodynamics with 
the proportionality of the entropy to the horizon 
area in general relativity. 
The procedure in Ref.~\cite{Jacobson:1995ab} was developed to 
more general extended gravitational 
theories~\cite{
Modified-gravity-EOS, Brustein:2009hy}.
On the other hand, 
to derive the corresponding gravitational field equation by using the 
formulation in Ref.~\cite{Jacobson:1995ab} in $f(R)$ gravity, 
a non-equilibrium thermodynamic treatment was 
proposed~\cite{Eling:2006aw}. 
It has  been recently demonstrated that 
one can acquire an equilibrium description of 
thermodynamics on the apparent horizon 
in the Friedmann-Lema\^{i}tre-Robertson-Walker (FLRW) space-time 
for modified gravity theories with the Lagrangian density $f(R, \phi, X)$,
where $X=-\left(1/2\right) g^{\mu\nu} {\nabla}_{\mu}\phi {\nabla}_{\nu}\phi$
is the kinetic term of a scalar field $\phi$ (${\nabla}_{\mu}$ is the 
covariant derivative operator associated with the metric tensor 
$g_{\mu \nu}$), 
by redefining an energy momentum tensor of ``dark'' components so that 
a local energy conservation law can be verified~\cite{Bamba-Geng-Tsujikawa}. 
The same consequence has been obtained in $f(R)$ gravity 
in the Palatini formalism~\cite{Bamba:2010kf}.

In this paper, 
we investigate both non-equilibrium and 
equilibrium descriptions of thermodynamics in $f(T)$ gravity. 
We also derive the conditions required by
the second law of thermodynamics.
We use units of $k_\mathrm{B} = c = \hbar = 1$ and denote the
gravitational constant $8 \pi G$ by 
${\kappa}^2 \equiv 8\pi/{M_{\mathrm{Pl}}}^2$ 
with the Planck mass of $M_{\mathrm{Pl}} = G^{-1/2} = 1.2 \times 10^{19}$GeV.

The paper is organized as follows. 
In Sec.\ II, we explain $f(T)$ gravity. 
In Sec.\ III, we investigate the non-equilibrium description of thermodynamics 
and explore the first and second laws of thermodynamics of 
the apparent horizon. 
In Sec.\ IV, we study the equilibrium description of thermodynamics. 
Finally, conclusions are given in Sec.\ V.

\section{$f(T)$ gravity}

In the teleparallelism, one uses 
orthonormal tetrad components 
$e_A (x^{\mu})$, 
where an index $A$ runs over $0, 1, 2, 3$ for the 
tangent space at each point $x^{\mu}$ of the manifold. 
Their relation to the metric $g^{\mu\nu}$ is given by 
\begin{equation}
g_{\mu\nu}=\eta_{A B} e^A_\mu e^B_\nu\,, 
\label{eq:2.1}
\end{equation}
where $\mu$ and $\nu$ are coordinate indices on the manifold 
and also run over $0, 1, 2, 3$, 
and $e_A^\mu$ forms the tangent vector of the manifold. 

The torsion $T^\rho_{\verb| |\mu\nu}$ and contorsion 
$K^{\mu\nu}_{\verb|  |\rho}$ tensors are defined by
\begin{eqnarray}
T^\rho_{\verb| |\mu\nu} \Eqn{\equiv} e^\rho_A 
\left( \partial_\mu e^A_\nu - \partial_\nu e^A_\mu \right)\,, 
\label{eq:2.2} \\ 
K^{\mu\nu}_{\verb|  |\rho} 
\Eqn{\equiv} 
-\frac{1}{2} 
\left(T^{\mu\nu}_{\verb|  |\rho} - T^{\nu \mu}_{\verb|  |\rho} - 
T_\rho^{\verb| |\mu\nu}\right)\,,
\label{eq:2.3}
\end{eqnarray}
respectively.
Instead of the Ricci scalar $R$ for the Lagrangian density 
in general relativity, 
the teleparallel Lagrangian density is described by the torsion scalar 
$T$, defined as 
\begin{equation}
T \equiv S_\rho^{\verb| |\mu\nu} T^\rho_{\verb| |\mu\nu}\,,
\label{eq:2.4}
\end{equation}
where
\begin{equation}
S_\rho^{\verb| |\mu\nu} \equiv \frac{1}{2}
\left(K^{\mu\nu}_{\verb|  |\rho}+\delta^\mu_\rho \ 
T^{\alpha \nu}_{\verb|  |\alpha}-\delta^\nu_\rho \ 
T^{\alpha \mu}_{\verb|  |\alpha}\right)\,. 
\label{eq:2.5}
\end{equation}
%
The modified teleparallel action for $f(T)$ gravity 
is given by~\cite{Linder:2010py} 
\begin{equation}
I= 
\int d^4x \abs{e} \left[ \frac{f(T)}{2{\kappa}^2} 
+{\mathcal{L}}_{\mathrm{M}} \right]\,,
\label{eq:2.6}
\end{equation}
where $\abs{e}= \det \left(e^A_\mu \right)=\sqrt{-g}$ and 
${\mathcal{L}}_{\mathrm{M}}$ is the matter Lagrangian. 
Varying the action in Eq.~(\ref{eq:2.6}) with respect to 
the vierbein 
vector field $e_A^\mu$, we obtain~\cite{Bengochea:2008gz} 
\begin{equation}
\frac{1}{e} \partial_\mu \left( eS_A^{\verb| |\mu\nu} \right) f^{\prime} 
-e_A^\lambda T^\rho_{\verb| |\mu \lambda} S_\rho^{\verb| |\nu\mu} 
f^{\prime} +S_A^{\verb| |\mu\nu} \partial_\mu \left(T\right) f^{\prime\prime} 
+\frac{1}{4} e_A^\nu f = \frac{{\kappa}^2}{2} e_A^\rho 
{T^{(\mathrm{M})}}_\rho^{\verb| |\nu}\,, 
\label{eq:2.7}
\end{equation}
where ${T^{(\mathrm{M})}}_\rho^{\verb| |\nu}$ 
is the contribution to the energy-momentum tensor from all 
perfect fluids of ordinary matter (radiation and non-relativistic matter).

We assume the four-dimensional flat 
Friedmann-Lema\^{i}tre-Robertson-Walker (FLRW) 
space-time with the metric, 
\begin{equation}
d s^2 = h_{\alpha \beta} d x^{\alpha} d x^{\beta}
+\tilde{r}^2 d \Omega^2\,,
\label{eq:2.8}
\end{equation}
where $\tilde{r}=a(t)r$, $x^0=t$ and $x^1=r$ with the two-dimensional 
metric $h_{\alpha \beta}={\rm diag}(1, -a^2(t))$. 
Here, $a(t)$ is the scale factor and 
$d \Omega^2$ is the metric of two-dimensional sphere with unit radius. 
In this space-time, 
$g_{\mu \nu}= \mathrm{diag} (1, -a^2, -a^2, -a^2)$ and 
the tetrad components $e^A_\mu = (1,a,a,a)$ 
yield the exact value of torsion scalar $T=-6H^2$, 
where $H=\dot{a}/a$ is the Hubble parameter 
and 
the dot denotes the time derivative of $\partial/\partial t$. 

In the flat FLRW background, 
it follows from Eq.~(\ref{eq:2.7}) that 
the modified Friedmann equations are given 
by~\cite{Bengochea:2008gz, Linder:2010py} 
\begin{eqnarray}
H^2 \Eqn{=} \frac{1}{6 F} 
\left( {\kappa}^2 \rho_{\mathrm{M}} -\frac{f}{2} \right)\,,
\label{eq:2.9} \\ 
\dot{H} 
\Eqn{=} 
-\frac{1}{ 4TF^{\prime} + 2F } 
\left( {\kappa}^2 P_{\mathrm{M}} -TF +\frac{f}{2} \right)\,, 
\label{eq:2.10}
\end{eqnarray}
where $F\equiv df/dT$, $F^{\prime}=dF/dT$,
and 
$\rho_{\mathrm{M}}$ and $P_{\mathrm{M}}$ are 
the energy density and pressure of all perfect fluids of generic matter, 
respectively. 
We note that the perfect fluid satisfies the continuity equation 
\begin{equation}
\dot{\rho}_{\mathrm{M}}+3H\left( \rho_{\mathrm{M}} + P_{\mathrm{M}} \right)
=0\,.
\label{eq:2.11}
\end{equation}
%

Before we proceed to the next section, 
we would summarize  our motivation of this paper. 
First, 
$f(T)$ gravity can realize the accelerated expansion of the universe 
(not only the late time cosmic acceleration but also inflation). 
Second, 
$f(T)$ gravity has  the second-order gravitational field equation in derivatives, 
similar to that in general relativity, 
whereas 
the gravitational field equation of $f(R)$ gravity is 
fourth-order in derivatives. 
Hence, it is important to investigate the theoretical  aspects 
in order to examine whether $f(T)$ gravity 
can be an alternative gravitational theory to general relativity. 
In this paper, we concentrate on the first and second laws of thermodynamics of 
$f(T)$ gravity. 

\section{Non-equilibrium description of thermodynamics in $f(T)$ gravity}

\subsection{Energy density and pressure of dark components}

Equations (\ref{eq:2.9}) and (\ref{eq:2.10}) can be rewritten as
\begin{eqnarray}  
H^2 
\Eqn{=}
\frac{\kappa^2}{3F} \left( \hat{\rho}_{\mathrm{DE}}+
\rho_{\mathrm{M}} \right)\,, 
\label{eq:3.1} \\
\dot{H}
\Eqn{=}
-\frac{\kappa^2}{2F} \left( \hat{\rho}_{\mathrm{DE}}+\hat{P}_{\mathrm{DE}}
+\rho_{\mathrm{M}}+P_{\mathrm{M}} \right)\,,
\label{eq:3.2} 
\end{eqnarray} 
where $\hat{\rho}_{\mathrm{DE}}$ and $\hat{P}_{\mathrm{DE}}$ are 
the energy density and pressure of ``dark'' components, given by 
\begin{eqnarray}  
\hat{\rho}_{\mathrm{DE}} 
\Eqn{\equiv} 
\frac{1}{2\kappa^2}\left( FT - f \right)\,, 
\label{eq:3.3} \\
\hat{P}_{\mathrm{DE}} 
\Eqn{\equiv} 
\frac{1}{2\kappa^2} 
\left[ -\left( FT - f \right) + 4H\dot{F}
\right]\,,
\label{eq:3.4}
\end{eqnarray} 
%
leading to 
\begin{eqnarray}
\dot{\hat{\rho}}_{\mathrm{DE}}+3H \left( 
\hat{\rho}_{\mathrm{DE}}+\hat{P}_{\mathrm{DE}} \right)
=-\frac{T}{2\kappa^2} \dot{F}\,. 
\label{eq:3.5}
\end{eqnarray} 
Here, 
a hat denotes quantities in the non-equilibrium 
description of thermodynamics 
as the standard continuity equation does not hold due to $\dot{F}\neq 0$ in Eq.~(\ref{eq:3.5}).

\subsection{First law of thermodynamics}

We investigate the thermodynamic property of $f(T)$ gravity. 
The relation
$h^{\alpha \beta} \partial_{\alpha} \tilde{r} \partial_{\beta} \tilde{r}=0$ 
determines the dynamical apparent horizon\footnote{
In Refs.~\cite{Zhou:2007pz,Wang:2005pk}, 
by using the  recent type Ia Supernovae data it has been shown that the accelerating universe enveloped by the apparent 
horizon satisfies the generalized second law of thermodynamics, 
whereas the accelerating universe inside the event horizon does not. Clearly, from the thermodynamics point of view,
the enveloping surface should be the apparent  horizon and not the event one in the accelerating universe.
}. 
In the flat FLRW space-time, the radius $\tilde{r}_A$ of 
the apparent horizon is given by 
\begin{equation}
\tilde{r}_A= \frac{1}{H}\,.
\label{eq:3.6}
\end{equation}
The time derivative of this relation gives to 
\begin{equation}
-\frac{d\tilde{r}_A}{\tilde{r}_A^3}
=\dot{H}H dt\,. 
\label{eq:3.7}
\end{equation}
Substituting Eq.~(\ref{eq:3.2}) into (\ref{eq:3.7}), we obtain
\begin{equation}
\frac{F}{4\pi G} d\tilde{r}_A=\tilde{r}_A^3 H
\left( \hat{\rho}_{\mathrm{t}}+\hat{P}_{\mathrm{t}}\right) dt\,,
\label{eq:3.8}
\end{equation}
where 
$\hat{\rho}_{\mathrm{t}} \equiv \hat{\rho}_{\mathrm{DE}}+
\rho_{\mathrm{M}}$ and $\hat{P}_{\mathrm{t}} \equiv \hat{P}_{\mathrm{DE}}+
P_{\mathrm{M}}$ 
are the total energy density and pressure of the universe, respectively. 

Note that in general relativity, the Bekenstein-Hawking horizon entropy is 
described by 
$S=A/\left(4G\right)$, where $A=4\pi \tilde{r}_A^2$ is the area of the 
apparent horizon~\cite{BCH-B-H}. 
In modified gravity theories including $f(R)$ gravity, 
a horizon entropy $\hat{S}$ associated with a Noether 
charge, called the Wald entropy $\hat{S}$~\cite{Wald entropy}, 
is expressed as 
$\hat{S}=A/\left(4G_{\rm eff}\right)$, where $G_{\mathrm{eff}}=G/f'$ 
with $f' = d f(R)/dR$ 
is the effective gravitational coupling 
in $f(R)$ gravity~\cite{Brustein:2007jj}. 
We remark that 
the Wald entropy in $f(R)$ gravity in both the metric~\cite{Wald entropy, 
Jacobson:1993vj}
 and 
Palatini~\cite{Vollick:2007fh} formalisms 
take the same form.

According to the study of the matter density perturbations 
in $f(T)$ gravity, 
the effective gravitational coupling in $f(T)$ gravity is 
given by $G_{\mathrm{eff}}=G/F$~\cite{Zheng:2010am}, 
similar to that in $f(R)$ gravity. 
Furthermore, 
recently, in Ref.~\cite{Miao:2011ki}, 
by using the Wald's Noether charge method~\cite{Wald entropy} 
and the related insights acquired 
in Refs.~\cite{Jacobson:1995ab, Brustein:2009hy, Eling:2006aw, 
Procedure-for-Thermodynamics}, 
it has been shown that 
when $F^{\prime}=dF(T)/dT = d^2 f(T)/d T^2$ is small, 
the entropy of black holes in $f(T)$ gravity 
is approximately equal to $FA/\left(4G\right)$. 
Hence, we take the Wald entropy in $f(T)$ gravity as 
\begin{equation}
\hat{S}=\frac{FA}{4G}\,. 
\label{eq:3.9}
\end{equation}
%
It is known that $f(T)$ gravity is not local Lorentz 
invariant~\cite{Local-Lorentz-invariance}, 
which indicates the existence of new degrees of freedom. 
However, at the background level no new degrees of
freedom are present, while at linear perturbation the new vector
degree of freedom only satisfies constraint equations~\cite{Local-Lorentz-invariance}.
It seems that these new degrees of freedom should not directly contribute to physical observables, especially in the early
universe~\cite{Wu}. Accordingly, we will assume that they do not contribute to the entropy in our study. 

Using Eqs.~(\ref{eq:3.8}) and (\ref{eq:3.9}), we find 
\begin{equation}
\frac{1}{2\pi \tilde{r}_A} d\hat{S}=4\pi \tilde{r}_A^3 H
\left( \hat{\rho}_{\mathrm{t}}+\hat{P}_{\mathrm{t}} \right) dt +
\frac{\tilde{r}_A}{2G} dF\,.
\label{eq:3.10}
\end{equation}
%

The associated temperature of the apparent horizon has 
the following Hawking temperature $T_{\mathrm{H}}$ 
%
\begin{equation}
T_{\mathrm{H}} 
=
\frac{|\kappa_{\mathrm{sg}}|}{2\pi}\,, 
\label{eq:3.11} 
\end{equation}
with $\kappa_{\mathrm{sg}}$ being the surface gravity, 
given by~\cite{Cai:2005ra} 
\begin{eqnarray}
\kappa_{\mathrm{sg}} 
\Eqn{=} 
\frac{1}{2\sqrt{-h}} \partial_\alpha 
\left( \sqrt{-h}h^{\alpha\beta} \partial_\beta \tilde{r} \right) 
\label{eq:3.12} \\
\Eqn{=} -\frac{1}{\tilde{r}_A}
\left( 1-\frac{\dot{\tilde{r}}_A}{2H\tilde{r}_A} \right)
=-\frac{\tilde{r}_A}{2} \left( 2H^2+\dot{H} \right) 
= 
-\frac{2\pi G}{3F} \tilde{r}_A 
\left( \hat{\rho}_{\mathrm{t}}-3\hat{P}_{\mathrm{t}} \right)\,, 
\label{eq:3.13}
\end{eqnarray}
where $h$ is the determinant of the metric $h_{\alpha\beta}$. 
{}From Eq.~(\ref{eq:3.13}), we see that 
$\kappa_{\mathrm{sg}} \le 0$ 
if 
the total equation of state (EoS) 
$w_{\mathrm{t}} \equiv \hat{P}_{\mathrm{t}}/\hat{\rho}_{\mathrm{t}}$ 
satisfies 
$w_{\mathrm{t}} \le 1/3$. 
Using Eqs.~(\ref{eq:3.11}) and (\ref{eq:3.13}), 
we have 
\begin{equation}
T_{\mathrm{H}}=\frac{1}{2\pi \tilde{r}_A}
\left( 1-\frac{\dot{\tilde{r}}_A}{2H\tilde{r}_A} \right)\,.
\label{eq:3.14}
\end{equation}
By multiplying the term $1-\dot{\tilde{r}}_A/(2H\tilde{r}_A)$ for 
Eq.~(\ref{eq:3.10}), we acquire 
%
\begin{equation}
T_{\mathrm{H}} d\hat{S} = 4\pi \tilde{r}_A^3 H \left(\hat{\rho}_{\mathrm{t}}+
\hat{P}_{\mathrm{t}} \right) dt 
-2\pi  \tilde{r}_A^2 \left(\hat{\rho}_{\mathrm{t}}+\hat{P}_{\mathrm{t}} 
\right) d \tilde{r}_A
+\frac{T_{\mathrm{H}}}{G}\pi \tilde{r}_A^2 dF\,. 
\label{eq:3.15}
\end{equation}

In general relativity, the Misner-Sharp energy~\cite{Misner:1964je}
is defined as $E \equiv \tilde{r}_A/\left(2G\right)$. 
Since $G_{\mathrm{eff}}=G/F$~\cite{Zheng:2010am} 
in $f(T)$ gravity, this may be extended to 
%
\begin{equation}
\hat{E}=\frac{\tilde{r}_AF}{2G}\,, 
\label{eq:3.16}
\end{equation}
similar to that in $f(R)$ gravity~\cite{Gong:2007md, Wu:2007se, Wu:2008ir} 
(for related works, see also~\cite{Sakai:2001gh}).
By combining Eqs.~(\ref{eq:3.6}) and (\ref{eq:3.16}), we find 
\begin{equation}
\hat{E}=
V \frac{3F H^2}{8\pi G}=V\hat{\rho}_{\mathrm{t}}\,,
\label{eq:3.17}
\end{equation}
where $V=4\pi \tilde{r}_A^3/3$ is the volume inside 
the apparent horizon. The last equality in Eq.~(\ref{eq:3.17}) 
means that $\hat{E}$ corresponds to the total intrinsic energy. 
%
It is clear from Eq.~(\ref{eq:3.17}) that $F>0$ so that 
$\hat{E} >0$. 
In this case, 
the effective gravitational coupling $G_\mathrm{eff} = G/F$  in $f(T)$ gravity 
becomes positive like $f(R)$ gravity~\cite{Sotiriou:2008rp}.
Note that 
this condition is required to ensure that
the graviton is not a ghost in the sense of quantum 
theory~\cite{Starobinsky:2007hu}.
%

Using Eqs.~(\ref{eq:2.11}) and (\ref{eq:3.5}), we find
\begin{equation}
d\hat{E} = -4\pi \tilde{r}_A^3 H \left(\hat{\rho}_{\mathrm{t}}
+\hat{P}_{\mathrm{t}} \right) dt 
+4\pi \tilde{r}_A^2  \hat{\rho}_{\mathrm{t}} 
d\tilde{r}_A+\frac{\tilde{r}_A }{2G} dF\,.
\label{eq:3.18}
\end{equation}
It follows from Eqs.~(\ref{eq:3.15}) and (\ref{eq:3.18}) that 
\begin{equation}
T_{\mathrm{H}} d\hat{S} = d\hat{E}+2\pi \tilde{r}_A^2 
\left(\hat{\rho}_{\mathrm{d}}+\rho_{\mathrm{f}}-\hat{P}_{\mathrm{d}}
-P_{\mathrm{f}}\right) d\tilde{r}_A 
+\frac{\tilde{r}_A}{2G} \left( 1+2\pi \tilde{r}_A T_{\mathrm{H}} \right) dF\,.
\label{eq:3.19}
\end{equation}
By introducing the work density~\cite{Cai:2006rs} 
\begin{eqnarray}
\hat{W} \Eqn{\equiv} 
-\frac{1}{2} \left( T^{(\mathrm{M})\alpha\beta}
h_{\alpha\beta} + \hat{T}^{(\mathrm{DE})\alpha\beta} h_{\alpha\beta} 
\right) 
\label{eq:3.20} \\
\Eqn{=} 
\frac12 \left(\hat{\rho}_{\mathrm{t}} 
-\hat{P}_{\mathrm{t}} \right)\,
\label{eq:3.21}
\end{eqnarray}
with $\hat{T}^{(\mathrm{DE})\alpha\beta}$ being 
the energy-momentum tensor of dark components, 
Eq.~(\ref{eq:3.19}) is rewritten to 
\begin{equation}
T_{\mathrm{H}} d\hat{S}=-d\hat{E}+\hat{W} dV
+\frac{\tilde{r}_A}{2G} 
\left( 1+2\pi \tilde{r}_A T_{\mathrm{H}} \right) dF\,, 
\label{eq:3.22}
\end{equation}
which can be described as 
\begin{equation}
T_{\mathrm{H}} d\hat{S}+T_{\mathrm{H}} d_{i}\hat{S}=-d\hat{E}+\hat{W} dV\,,
\label{eq:3.23}
\end{equation}
where 
\begin{eqnarray}
d_{i}\hat{S} \Eqn{=} -\frac{1}{T_{\mathrm{H}}} \frac{\tilde{r}_A}{2G}
\left( 1+2\pi \tilde{r}_A T_{\mathrm{H}} \right) dF
=-\left( \frac{\hat{E}}{T_{\mathrm{H}}}+\hat{S} \right) 
\frac{dF}{F} 
\label{eq:3.24} \\
\Eqn{=} \frac{6\pi}{G} \frac{8HT+\dot{T}}
{T\left(4HT+\dot{T}\right)} dF\,. 
\label{eq:3.25}
\end{eqnarray}
The additional term 
$d_{i}\hat{S}$ in Eq.~(\ref{eq:3.23}) can be interpreted as 
an entropy production term in the non-equilibrium thermodynamics. 
In $f(T)$ gravity, 
$d_{i}\hat{S}$ in Eq.~(\ref{eq:3.24}) in non-vanishing  
due to $d F \neq 0$ unless $f(T)=T$ which gives rise to  $F=1$ and 
$d_{i}\hat{S}=0$.
As a result,
the first-law of equilibrium thermodynamics holds. 

\subsection{Second law of thermodynamics}

Normally, in  the cosmological FLRW fluid 
dynamics the entropy is simply the fluid-entropy 
current and has little to do with the horizon entropy. 
 In the flat FLRW space-time, 
the Bekenstein-Hawking horizon entropy of the apparent horizon 
is expressed as 
$S=A/\left(4G\right) = \pi/\left( G H^2 \right) \propto H^{-2}$, 
where in deriving the first equality 
we have used $A=4\pi \tilde{r}_A^2$ and 
Eq.~(\ref{eq:3.6}). 
On the other hand, for modified gravitational theories including 
$f(R)$  and $f(T)$, 
there can exist the phantom phase in which $\dot{H} > 0$. 
In such phase, 
$\dot{S} = -2\left[ \pi/\left( G H^3 \right) \right] \dot{H} < 0$ and hence,
the second law of thermodynamics in terms of the horizon entropy 
is not satisfied. Thus, this seems to imply that 
such modified gravitational theories with the phantom phase 
cannot be an alternative gravitational theory to general relativity.
However, 
it is not the case. 
If we investigate the entropy of the total energy of the 
horizon, i.e., the sum of the horizon entropy and 
the entropy of ordinary perfect fluids of generic matter, 
the total entropy always increases in time 
and the second law of thermodynamics can be met, 
as explicitly shown in $f(R)$ gravity in Ref.~\cite{Bamba:2010kf}. 
In this subsection, we examine this point in $f(T)$ gravity. 

To explore the second law of thermodynamics in $f(T)$ gravity, 
we start with the Gibbs equation in terms of all matter and energy fluid, 
given by 
\begin{equation}
T_{\mathrm{H}} d\hat{S}_{\mathrm{t}} = d\left( \hat{\rho}_{\mathrm{t}} V \right) +\hat{P}_{\mathrm{t}} dV 
= V d\hat{\rho}_{\mathrm{t}} + \left( \hat{\rho}_{\mathrm{t}} + \hat{P}_{\mathrm{t}} \right) dV\,, 
\label{eq:3.26}
\end{equation}
where $T_{\mathrm{H}}$ and $\hat{S}_{\mathrm{t}}$ denote the temperature and 
entropy of total energy inside the horizon, respectively. 
Here, we have assumed the same temperature between the outside 
and inside of the apparent horizon. 
To obey the second law of thermodynamics in $f(T)$ gravity, 
we require that~\cite{Wu:2008ir}
\begin{equation}
\Xi \equiv 
\frac{d\hat{S}}{dt} + \frac{d\left( d_i \hat{S} \right)}{dt} 
+ \frac{d\hat{S}_{\mathrm{t}}}{dt} \geq 0\,. 
\label{eq:3.27}
\end{equation}
By using Eqs.~(\ref{eq:3.1}), (\ref{eq:3.23}) and (\ref{eq:3.26}), we obtain 
\begin{equation}
\Xi = -\frac{3}{4G} \frac{\dot{T}^2 F}{T^3}\,. 
\label{eq:3.28}
\end{equation}
Since $-T^3 = 216 H^6 >0$, 
we see that 
the condition of $\Xi\geq 0$ becomes 
\begin{equation}
J \equiv \dot{T}^2 F = 144H^2 \dot{H}^2 F \geq 0\,, 
\label{eq:3.29}
\end{equation}
which is always met because 
$F>0$ in order that $\hat{E} >0$. 
Hence, the second law of thermodynamics in $f(T)$ gravity can be satisfied. 
It is clear from Eq.~(\ref{eq:3.29}) that 
$J \geq 0$ irrespective of the sign of $\dot{H}$. 
This result is also consistent 
with a phantom model with ordinary thermodynamics~\cite{Nojiri:2005sr}. 
Investigations on entropy in phantom models have also been executed~\cite{Entropy-in-phantom-models}. 

We remark that we have used the physical temperature 
as the temperature of the apparent horizon, $i.e.$,
the Hawking temperature, given by 
$T_{\mathrm{H}} = |\kappa_{\mathrm{sg}}|/\left(2\pi\right)$ in Eq.~(\ref{eq:3.11}). 
This temperature  clearly depends on
the energy momentum tensor of the dark components 
coming from $f(T)$ gravity as demonstrated in, $e.g.,$ Eqs.~(\ref{eq:3.12}) and (\ref{eq:3.13}). 
On the other hand, in a cosmological setup 
the temperature of matter species, such as about $2.73$K of the CMB 
photons is determined in a standard way.
In our discussions on the second law of thermodynamics
in $f(T)$ gravity, 
we have concentrated on the case in which 
the temperature of the universe inside the 
horizon is equal to that of the apparent horizon. 
In other words, the temperature of the apparent horizon is assumed to be 
the same as the temperature of matter species including that of the CMB photons. 

\section{Equilibrium description of thermodynamics in $f(T)$ gravity}

Since there is a non-equilibrium entropy production term $d_i \hat{S}$,
the right-hand side (r.h.s.) of Eq.~(\ref{eq:3.5}) does not vanish, 
i.e., the standard continuity equation for  
$\hat{\rho}_{\mathrm{DE}}$ and $\hat{P}_{\mathrm{DE}}$ defined in 
Eqs.~(\ref{eq:3.3}) and (\ref{eq:3.4}) 
does not hold. 
In this section, 
it is demonstrated that 
by redefining 
the energy density and pressure of dark components to meet the 
continuity equation, 
there can be no extra entropy production term. 
We refer to this as the equilibrium description 
in $f(T)$ gravity.

\subsection{Energy density and pressure of dark components}

By comparing the 
gravitational field equations (\ref{eq:2.9}) and 
(\ref{eq:2.10}) with the ordinary ones in general relativity: 
\begin{eqnarray}
H^2 \Eqn{=} \frac{{\kappa}^2}{3} \left(\rho_{\mathrm{M}}+\rho_{\mathrm{DE}} 
\right)\,, 
\label{eq:4.1} \\ 
%
%
\dot{H}
\Eqn{=} -\frac{{\kappa}^2}{2} \left(\rho_{\mathrm{M}} + P_{\mathrm{M}} + 
\rho_{\mathrm{DE}} + P_{\mathrm{DE}} \right)\,,
\label{eq:4.2} 
\end{eqnarray}
the energy density and pressure of dark components 
can be rewritten as
%
\begin{eqnarray}
\rho_{\mathrm{DE}} 
\Eqn{=} 
\frac{1}{2{\kappa}^2} 
\left( -T -f +2TF \right)\,, 
\label{eq:4.3} \\ 
P_{\mathrm{DE}} 
\Eqn{=} 
-\frac{1}{2{\kappa}^2} 
\left[ 
4\left(1 -F -2TF^{\prime} \right) \dot{H} 
+\left( -T -f +2TF \right)
\right]\,, 
\label{eq:4.4} 
\end{eqnarray}
which clearly satisfy the standard continuity equation, i.e., 
%
\begin{equation} 
\dot{\rho}_{\mathrm{DE}}+3H \left( 
\rho_{\mathrm{DE}} + P_{\mathrm{DE}}
\right)  
= 0\,. 
\label{eq:4.5}
\end{equation} 
%

\subsection{First law of thermodynamics}

In the representation of Eqs.~(\ref{eq:4.1}) and (\ref{eq:4.2}), 
Eq.~(\ref{eq:3.8}) becomes 
\begin{equation} 
\frac{1}{4\pi G} d\tilde{r}_A=\tilde{r}_A^3 H
\left( \rho_{\mathrm{t}}+P_{\mathrm{t}} \right) dt\,, 
\label{eq:4.6}
\end{equation} 
with 
$\rho_{\mathrm{t}} \equiv \rho_{\mathrm{DE}}+
\rho_{\mathrm{M}}$ and $P_{\mathrm{t}} \equiv P_{\mathrm{DE}}+P_{\mathrm{M}}$. 
By introducing the horizon entropy $S=A/(4G)$
%
%
and using Eq.~(\ref{eq:4.6}), we have 
\begin{equation} 
\frac{1}{2\pi \tilde{r}_A} dS=4\pi \tilde{r}_A^3 H
\left( \rho_{\mathrm{t}}+P_{\mathrm{t}} \right) dt \,.
\label{eq:4.8}
\end{equation} 
{}From the horizon temperature in Eq.~(\ref{eq:3.14}) and Eq.~(\ref{eq:4.8}), 
we find 
\begin{equation} 
T_{\mathrm{H}} dS = 4\pi \tilde{r}_A^3 H \left(\rho_{\mathrm{t}}+P_{\mathrm{t}} \right) dt 
-2\pi  \tilde{r}_A^2 \left(\rho_{\mathrm{t}}+P_{\mathrm{t}} \right) 
d\tilde{r}_A\,.
\label{eq:4.9}
\end{equation} 

By defining the Misner-Sharp energy as 
\begin{equation} 
E=\frac{\tilde{r}_A}{2G}=
V\rho_{\mathrm{t}}\,,
\label{eq:4.10}
\end{equation} 
we get 
\begin{equation} 
dE=-4\pi \tilde{r}_A^3 H \left(\rho_{\mathrm{t}}+P_{\mathrm{t}} \right) dt 
+4\pi \tilde{r}_A^2 \rho_{\mathrm{t}} d\tilde{r}_A\,,
\label{eq:4.11}
\end{equation} 
where there does not exist any additional term proportional to $dF$ 
on the r.h.s. due to the continuity equation (\ref{eq:4.6}). From
Eqs.~(\ref{eq:4.9}) and (\ref{eq:4.11}), we obtain 
the following equation corresponding to the first law of 
equilibrium thermodynamics: 
\begin{equation} 
T_{\mathrm{H}} dS=-dE+W dV\,
\label{eq:4.12}
\end{equation} 
with the work density $W$
given by 
\begin{equation} 
W=\frac12 \left( \rho_{\mathrm{t}}-P_{\mathrm{t}} \right)\,.
\label{eq:4.13}
\end{equation} 
Thus, 
by redefining $\rho_{\mathrm{DE}}$ and 
$P_{\mathrm{DE}}$ so that the continuity equation (\ref{eq:4.5}) 
can be met, 
we can realize 
an equilibrium description of thermodynamics. 

Furthermore, 
it follows from Eqs.~(\ref{eq:4.1}), (\ref{eq:4.2}) and (\ref{eq:4.8}) 
that 
\begin{equation}
\dot{S} = 
8\pi^2 H \tilde{r}_A^4 \left(\rho_{\mathrm{t}}+P_{\mathrm{t}}\right) 
= \frac{6\pi}{G} \frac{\dot{T}}{T^2}\,. 
\label{eq:4.14}
\end{equation}
Since $\dot{S} \propto \dot{T}/T^2 \propto -\dot{H}/H^3$, 
the horizon entropy increases in the expanding universe 
as long as the null energy condition 
$\rho_{\mathrm{t}}+P_{\mathrm{t}} 
\ge 0$ is satisfied, in which 
$\dot{H} \leq 0$. 

There are two main reasons why we can obtain the equilibrium description of 
thermodynamics. 
One is that we can redefine the energy density and pressure of dark components 
so that the standard continuity equation can be satisfied. 
The other is that the horizon entropy $S$ is proportional to 
the horizon area $A$ 
in $S=A/(4G)$ 
as 
general relativity\footnote{
There also exist studies on 
the expression of the horizon entropy in the four-dimensional 
modified gravity~\cite{Wang:2005bi} and the quantum logarithmic correction to 
it in cosmological 
settings~\cite{Cai:2008ys}.
}.

The relation between 
the horizon entropy $S$ in the equilibrium description 
and $\hat{S}$ in the non-equilibrium description 
can be described as~\cite{Bamba-Geng-Tsujikawa} 
\begin{equation}
dS = d\hat{S} + d_i \hat{S} 
+\frac{\tilde{r}_A}{2GT_{\mathrm{H}}} dF 
-\frac{2\pi \left(1-F\right)}{G}
\frac{\dot{H}}{H^3}\, dt.
\label{eq:3.A-1}
\end{equation}
By using the relations (\ref{eq:3.25}) and (\ref{eq:4.8}), 
Eq.~(\ref{eq:3.A-1}) is rewritten to the following form: 
\begin{equation}
dS=\frac{1}{F} d\hat{S}+\frac{1}{F}
\frac{2H^2+\dot{H}}{4H^2+\dot{H}}\,d_i \hat{S}\,, 
\label{eq:3.A-2}
\end{equation}
where $d_i \hat{S}$ is given by Eq.~(\ref{eq:3.25}). 
The difference between $S$ and $\hat{S}$ appears in $f(T)$ gravity due to 
$dF \neq 0$, 
whereas 
$S = \hat{S}$ in the theory of $f(T)=T$ because $F=1$. 
Eq.~(\ref{eq:3.A-2}) implies 
that $dS$ in the equilibrium framework 
includes the information of 
$d\hat{S}$ as well as 
$d_i \hat{S}$ in the non-equilibrium framework.

In the flat FLRW space-time, 
since the Bekenstein-Hawking entropy 
behaves as 
$S \propto H^{-2}$ regardless of
gravity theories, 
$S$ increases 
if $H$ decreases,
whereas when $H$ becomes large, $S$ becomes small 
(similar to that in the case of superinflation), 
i.e., $S$ increases and decreases for 
$w_{\mathrm{t}} \equiv P_{\mathrm{t}}/\rho_{\mathrm{t}} > -1$ 
($\dot{H}<0$)
and 
$w_{\mathrm{t}}<-1$ ($\dot{H}>0$), respectively. 
That is, such a property is equivalent to the standard general relativistic 
picture with the energy density $\rho_{\mathrm{DE}}$ in Eq.~(\ref{eq:4.3}) and 
the pressure $P_{\mathrm{DE}}$ in Eq.~(\ref{eq:4.4}) of dark components. 

On the other hand, 
the Wald entropy 
evolves as $\hat{S} \propto FH^{-2}$ 
with involving the information of theories. 
For instance, 
in a model of $f(T)= T + \alpha T^n$, where 
$\alpha$ and $n$ are constants, 
$\hat{S} \propto H^{2(n-2)}$ and hence 
$\hat{S}$ becomes large apart from $n=2$ 
because $H$ increases (decreases) for $n>2$ ($n<2$). 
Therefore, the behavior of the Wald entropy 
is different from that of the Bekenstein-Hawking entropy. 
The entropy production term $d_{i}\hat{S}$ 
in the non-equilibrium description 
presents a relation to the equilibrium description 
as given in Eq.~(\ref{eq:3.A-2}), 
i.e., $dS$ consists of $d\hat{S}$ and $d_{i}\hat{S}$. 
As a result, 
the equilibrium framework is more transparent than 
the non-equilibrium one because 
it gives not only 
the general relativistic analogue 
of the horizon entropy independent of gravitational theories 
but also the more profound connection of the non-equilibrium thermodynamics 
with the standard equilibrium one. 

It is interesting to mention 
that there exists the difference between 
the EoS of dark components in the two
descriptions.
{}From Eqs.~(\ref{eq:3.3}), (\ref{eq:3.4}), 
(\ref{eq:4.3}) and (\ref{eq:4.4}), 
we find that 
\begin{eqnarray} 
\hat{w}_{\mathrm{DE}} = {\hat{P}_{\mathrm{DE}} \over \hat{\rho}_{\mathrm{DE}}}
\Eqn{=} -1 + \frac{4H\dot{F}}{FT - f}\,,
\label{eq:3.A-4} \\ 
w_{\mathrm{DE}} = {P_{\mathrm{DE}} \over  \rho_{\mathrm{DE}}}
\Eqn{=} -1 -\frac{4\left(1 -F -2TF^{\prime} \right) \dot{H}}
{-T -f +2TF}\,, 
\label{eq:3.A-5}
\end{eqnarray} 
in the non-equilibrium and equilibrium descriptions, respectively.
It is easy to see that in general 
$\hat{w}_{\mathrm{DE}} \neq w_{\mathrm{DE}}$ except 
the theory of $f(T)=T$.
Consequently, 
in order to compare the theoretical prediction in terms of the EoS of dark 
components in $f(T)$ gravity with the observations, 
the expression of the EoS of dark components in the
non-equilibrium description as well as that in the equilibrium one are necessary. 
This is a meaningful physical implication of the results from the physics 
perspective obtained by exploring both non-equilibrium and equilibrium 
descriptions. 

\subsection{Second law of thermodynamics}

To examine the second law of thermodynamics in the equilibrium description, 
we write the Gibbs equation in terms of all matter and energy fluid as 
\begin{equation}
T_{\mathrm{H}} dS_{\mathrm{t}} = d\left( \rho_{\mathrm{t}} V \right) +P_{\mathrm{t}} dV 
= V d\rho_{\mathrm{t}} + \left( \rho_{\mathrm{t}} +P_{\mathrm{t}} 
\right) dV\,. 
\label{eq:4.A001}
\end{equation}
The second law of thermodynamics can be described by 
\begin{equation}
\frac{dS_{\mathrm{sum}}}{dt} 
\equiv 
\frac{dS}{dt} + \frac{dS_{\mathrm{t}}}{dt} 
\geq 0\,,
\label{eq:4.A002}
\end{equation}
where $S_{\mathrm{sum}} \equiv S + S_{\mathrm{t}}$.
Consequently, we obtain 
\begin{equation}
\frac{dS_{\mathrm{sum}}}{dt} 
= 
-\frac{6\pi}{G}\left( \frac{\dot{T}}{T} \right)^2 \frac{1}{4HT + \dot{T}}\,,
\label{eq:4.A003}
\end{equation}
by using $V=4\pi \tilde{r}_A^3/3$, 
and 
Eqs.~(\ref{eq:3.14}), (\ref{eq:4.2}) and (\ref{eq:4.14}). 
Hence, the relation (\ref{eq:4.A002}) with Eq.~(\ref{eq:4.A003}) 
leads to the condition 
\begin{equation}
Y \equiv -\left( 4HT + \dot{T} \right) 
= 12H \left( 2H^2 + \dot{H} \right) 
\geq 0\,. 
\label{eq:4.A004}
\end{equation}
%
In the flat FLRW expanding background ($H>0$), 
the second law of thermodynamics can be satisfied 
as long as the quantity $\left( 2H^2 + \dot{H} \right)$ 
is positive or equal to zero. 
It is interesting to note that in $f(R)$ gravity,
$R=6\left( 2H^2 + \dot{H} \right)$~\cite{Bamba:2010kf} for the flat FLRW space-time
and the condition in Eq.~(\ref{eq:4.A004}) clearly holds as $R \geq 0$.
By analogy with $f(R)$ gravity, we can extend the condition, i.e., 
$Y\geq 0$,  to $f(T)$ gravity since $Y$ involves only 
$H$ and $\dot{H}$ and is related to the scalar curvature in general relativity.
Thus, we have acquired a unified insight between non-equilibrium 
and equilibrium descriptions of thermodynamics. 
It should be cautioned that 
this result can be shown explicitly only for the same 
temperature of the universe outside and inside the apparent horizon~\cite{Gong:2006ma}.

\section{Conclusions}

We have investigated 
the first and second laws of 
thermodynamics of the apparent horizon in $f(T)$ gravity 
with both non-equilibrium and equilibrium descriptions. 
We have found the same dual equilibrium/non-equilibrium formulation for $f(T)$ 
as for $f(R)$ gravity. 
We have demonstrated 
that the second law of thermodynamics can be satisfied 
in the non-equilibrium and equilibrium frameworks
if the temperature of the universe inside the 
horizon is equal to that of the apparent horizon. 
It has been shown that 
in the non-equilibrium framework, 
the second law of thermodynamics can be met 
regardless of the sign of the time derivative of 
the Hubble parameter, 
whereas 
in the equilibrium framework, 
the second law of thermodynamics can be verified 
by analogy with 
the same non-negative quantity related to the scalar curvature 
in general relativity is positive or equal to zero 
in the expanding cosmological background. 

Finally, we emphasize that our result of
the second law of thermodynamics in $f(T)$ gravity 
is nontrivial and meaningful.
We believe that any successful alternative 
gravitational theory to general relativity should obey
this law. 
If the  law is violated in certain
universes in a model,  it is more likely to be due to an incorrect generalization
of the second law or some inherent inconsistency of the model itself. For the latter,
the model should be abandoned.
Furthermore, 
it is important to note that
the same temperature  inside and on the apparent horizon 
is  a working hypothesis as  it may not be so generally.

\section*{Acknowledgments}

We would like to thank Professor Shinji Tsujikawa for his collaboration 
in our previous work~\cite{Bamba-Geng-Tsujikawa}. 
The work is supported in part by 
the National Science Council of R.O.C. under
Grant \#: 
NSC-98-2112-M-007-008-MY3
and 
National Tsing Hua University under the Boost Program \#: 
99N2539E1.


\end{document}